\documentclass[aps,pra,twocolumn,superscriptaddress,byrevtex,floatfix]{revtex4}

\usepackage{graphicx}% Include figure files
\usepackage{dcolumn}% Align table columns on decimal point
\usepackage{bm}% bold math

\begin{document}

%\preprint{Version 5.0}

\title{Nearly strain-free heteroepitaxial system for fundamental studies
of pulsed laser deposition: EuTiO$_3$ on SrTiO$_3$}

\author{Huan-Hua Wang}
\altaffiliation{Current Address: Department of Physics and Astronomy, University of Georgia, Athens, GA 30602-2451}
\affiliation{School of Applied and Engineering Physics, Cornell University, Ithaca, NY 14853}
\affiliation{Cornell Center for Materials Research, Cornell University, Ithaca, NY 14853}
\author{Aaron Fleet}
\affiliation{School of Applied and Engineering Physics, Cornell University, Ithaca, NY 14853}
\affiliation{Cornell Center for Materials Research, Cornell University, Ithaca, NY 14853}
\author{Darren Dale}
\affiliation{Department of Materials Science and Engineering, Cornell University, Ithaca, NY 14853}
\affiliation{Cornell Center for Materials Research, Cornell University, Ithaca, NY 14853}
\author{Yuri Suzuki}
\altaffiliation{Current Address: Department of Materials Science and Engineering, University of California at Berkeley, Berkeley, CA 94720-1760}
\affiliation{Department of Materials Science and Engineering, Cornell University, Ithaca, NY 14853}
\affiliation{Cornell Center for Materials Research, Cornell University, Ithaca, NY 14853}
\author{J.D. Brock}
\affiliation{School of Applied and Engineering Physics, Cornell University, Ithaca, NY 14853}
\affiliation{Cornell Center for Materials Research, Cornell University, Ithaca, NY 14853}

\date{\today}

\begin{abstract}
High quality
epitaxial thin-films of EuTiO$_3$ have been grown
on the (001)
surface of SrTiO$_3$ using 
pulsed-laser deposition.
{\it In situ}\/ x-ray reflectivity measurements
reveal that the growth is two-dimensional and enable 
real-time monitoring of the film thickness and roughness during growth.
The film thickness, surface mosaic, surface roughness, and strain were characterized
in detail using
{\it ex situ}\/ x-ray diffraction.
The thickness and composition were confirmed with Rutherford Back-Scattering.
%The roughness and surface morphology of the films were also characterized with
%{\it ex situ} Atomic Force Microscopy.
The EuTiO$_3$ thin-films grow two-dimensionally, epitaxially,
pseudomorphically,
with no measurable in-plane lattice
mismatch.
%The out-of-plane lattice parameter is mismatched by approximately 0.4\%.
\end{abstract}

\maketitle

\section*{Introduction}

The growth of thin-films pervades modern electronic device fabrication.
Indeed, the physical properties of thin-films are greatly influenced by their quality.
And, the quality of a film is largely determined during its growth.
Hence, optimizing growth processes is a significant theme of current research.
Recently, thin-films of the transition metal oxides have been studied intensively due to their ferroelectric,
ferromagnetic, and superconducting properties\cite{imada_98}.
Pulsed-laser deposition (PLD) is widely used to deposit thin-films of these and other materials 
with non-trivial stoichiometries\cite{chrisey_94, willmott_00, lowndes_96}.
In PLD an intense pulse of laser light creates a ``plume'' of material by ablating it from a target.
This plume exists for a few microseconds and consists of both 
ionic and neutral species.
The energetic character of the plume distinguishes PLD from equilibrium growth techniques such as
molecular-beam epitaxy (MBE) and chemical vapor deposition (CVD),
while the enormous instantaneous flux distinguishes PLD from other energetic growth techniques such 
as sputter deposition or ion-beam assisted deposition (IBAD).
While it is well-known that PLD 
can produce high quality epitaxial thin-films, our fundamental (atomic level) understanding of 
the PLD growth process remains incomplete.
%The long-term goal of this research program is to 
%determine the atomic mechanisms that are operant in PLD\@. 

Our long-term interest is identifying and studying the atomic scale, possibly
energetic
({\em i.e.}\/, non-thermal), mechanisms
operant in PLD growth.
Many factors, including differences 
in the symmetry of the substrate and thin-film lattices,
a mismatch of the lattice constants,
the partial pressure of oxygen, and the substrate temperature
are known to strongly influence the growth process.
Therefore,
to isolate novel growth mechanisms,
we require a nearly ideal thin-film/substrate combination that has consistent
symmetry and that eliminates lattice mismatch.
The objective of this study is to develop such a material system.

We have found that
PLD of EuTiO$_3$ (ETO) thin-films on SrTiO$_3$ (STO) substrates satisfies these requirements.
ETO and STO both have the cubic perovskite crystal structure at room temperature and both have the lattice 
constant $3.905\:$\AA\cite{holzapfel_66,mccarthy_69,katsufuji_01}.
The crystal structure of ETO is truly centrosymmetric, while STO exhibits only a 
tiny distortion from centrosymmetry\cite{ravel_95}.
In addition, neither material goes through a structural phase 
transition between $-220^\circ$C and typical PLD growth temperatures of $600 - 800^\circ$C\cite{galasso_69}.
Previous X-ray 
studies in the literature indicate that oxygen-deficient solid solutions of EuTiO$_x$ ($2.5 < x < 3$) 
maintain the perovskite structure
for the entire range of $x$ and do not report any change in the lattice constant
\cite{chien_74,mccarthy_69b}.

\section*{Experimental Details}
These ETO/STO PLD studies are the first thin-film growth experiments performed in the 
new thin-film growth/time-resolved x-ray diffraction facility in the G3 experimental station at 
the Cornell High Energy Synchrotron Source (CHESS),
which utilizes x-rays generated by the Cornell Electron Storage Ring (CESR)\@.
The G3 experimental station combines 
a state-of-the-art ultra-high-flux synchrotron x-ray scattering facility with a fully featured PLD 
growth system, with the goal of performing time-resolved structural measurements of thin-film 
growth.
In these experiments, the (001) STO substrate sits at the origin of an x-ray diffractometer whose motions are integral components of the PLD growth chamber.
%The $\sim100$ picosecond x-ray pulses from the storage ring will ultimately enable us to perform time-resolved x-ray 
%measurements on (sub-)nanosecond time-scales.

\begin{figure}[p]
\noindent
\begin{center}
\includegraphics[width=\linewidth]{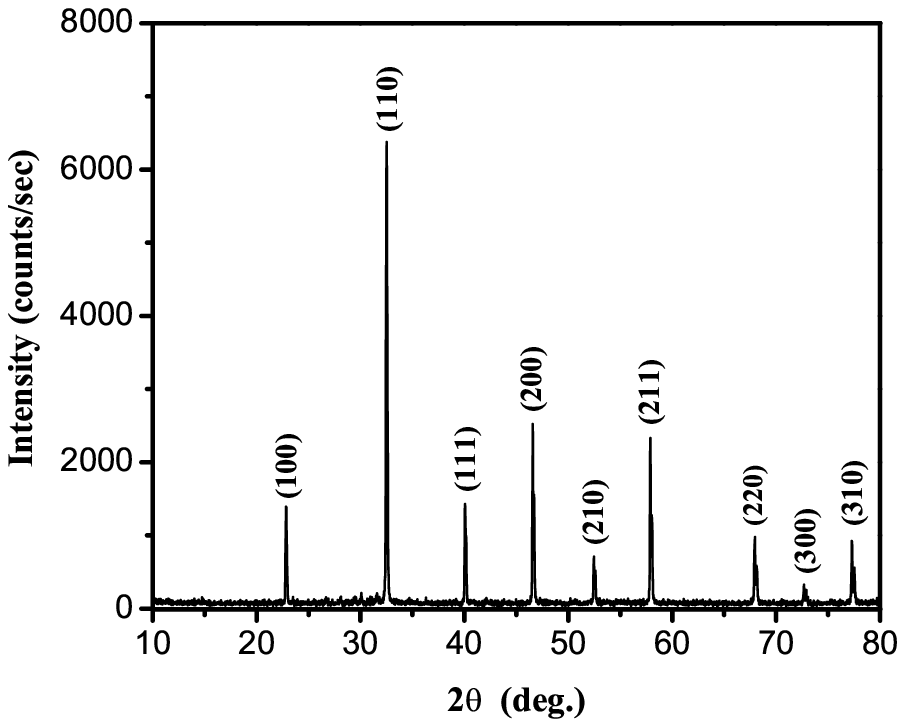}
\end{center}
\caption{XRD scan of EuTiO$_3$ powder sample before being sintered into a pellet for use as a PLD target.
All of the visible peaks have been indexed to the cubic perovskite structure.
Note the absence of any peaks associated with a minority phase.}
\label{fig:powder}
\end{figure}

The ETO PLD target was synthesized by a solid-state reaction.
Commercial Eu$_2$O$_3$ 
and TiO$_2$ powders were mixed and compressed into a pellet.
The pellet was then sintered at 
1000$^\circ$C for two days to obtain Eu$_2$Ti$_2$O$_7$\@.
The pellet was then reground into a powder and reduced in hydrogen gas at 1180$^\circ$C into EuTiO$_3$\@.
The powder was characterized using standard X-ray diffraction (XRD) techniques.
The powder needed to be reground and reduced several times in order to obtain pure EuTiO$_3$\@.
The final powder diffraction profile, which does not exhibit any indication of a minority phase,
is shown in Figure [\ref{fig:powder}]\@. 
The measured lattice constant is $3.896\:$\AA, quite close to the previously
reported value\cite{brous_53}.
Finally, the purified powder was pressed into a pellet and vacuum sintered.

The substrates used in this study were commercially cut and polished
(001) single-crystal wafers of SrTiO$_3$\@.
Each substrate was fixed to the stainless steel heater with silver paste as received
from the supplier with no additional surface preparation.
The target-substrate distance was 7 cm.
The heater temperature was measured by a 
thermocouple inserted into the heater.
The substrate temperature is roughly $100^\circ$C (at $650^\circ$C) lower than 
the interior of the heater.
Depositions were performed at heater temperatures ranging from $50^\circ$C
to $650^\circ$C\@.

The plume is created by a pulsed excimer laser (Lambda Physik, 248 nm) focused on the 
(rotating) target with a fluence of 2 J/cm$^2$ and a repetition rate that ranged from 0.1 to 10 Hz, in a 
background of $7 \times 10^{-6}$ Torr of O$_2$\@.

{\it In situ}  x-ray measurements were performed in real-time to monitor the growth process.  
Specifically, the time dependence of the intensity of the x-rays scattered at the anti-Bragg condition
(equivalent to the out-of-phase condition in RHEED or LEED) exhibits intensity oscillations 
as the film grows.
These oscillations can be caused both by variations in the roughness of the growth surface 
during layer-by-layer growth (as in RHEED studies) and by interference between x-rays reflected
by the film surface and the film/substrate interface (Kiessig fringes)\cite{kiessig_31, als-nielsen}.
This latter effect is only present in 
heteroepitaxy and is not usually observable in RHEED studies;
however, it is very powerful as a time-resolved probe of film thickness.
The intensities of the Eu L$_{2,3}$ emission lines were also monitored,
providing an independent {\it in situ}\/ measurement 
of the thickness of the growing ETO film \cite{headrick_96, headrick_98, woll_99}.

After deposition, the film was cooled down to room temperature under the same oxygen 
pressure.
The as-grown films are optically transparent.
The thickness, surface roughness and 
crystal quality were characterized {\it ex situ}  with a standard rotating anode Cu K$_{\alpha}$
x-ray source.
%The surface morphology of the samples was imaged and characterized using
%Atomic Force Microscopy (AFM)\@.
Rutherford Back Scattering (RBS) was used to characterize the film thickness and 
composition.

\section*{Experimental Results}
\begin{figure}[p]
\noindent
\begin{center}
\includegraphics[width=\linewidth]{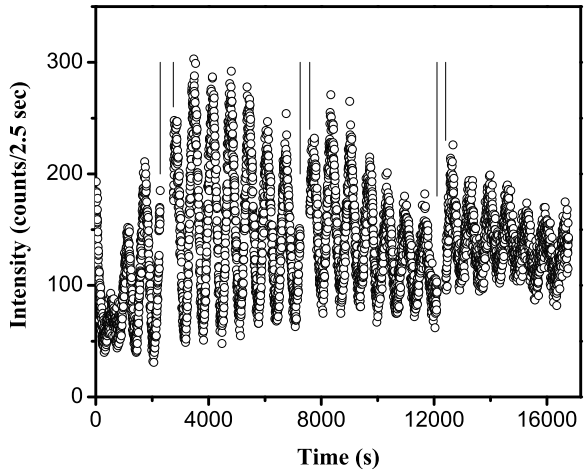}
\end{center}
\caption{Oscillations of the intensity of specularly reflected x-rays at the anti-Bragg position
during PLD growth of EuTiO$_3$ on SrTiO$_3$ (001).
The oscillations are Kiessig fringes.
Each oscillation corresponds roughly to adding two layers of EuTiO$_3$\@.
Gaps in the time series correspond to CESR injections.}
\label{fig:oscillations}
\end{figure}

Figure [\ref{fig:oscillations}] shows an example of the time-dependent x-ray intensity
oscillations observed during PLD growth of ETO on STO (001)
at $650^\circ$C with a laser repetition rate of 0.1 Hz.
The deposition was stopped after 24 oscillation cycles.
(Gaps in the time series occur during CESR injections.)
In these measurements, the abrupt changes in intensity associated with each laser 
pulse (for example, as reported by G. Eres,{\it  et al.}\cite{eres_02}) are not visible
for two reasons.
First, our pulses of material are smaller.
During this growth, roughly 40 laser pulses were required to add 
a unit cell thick layer of ETO to the film.
Second, the temporary x-ray optics available
were not able to deliver the full x-ray beam, limiting the signal rate.
Nevertheless,
the observed intensity oscillations are clear evidence of
two-dimensional growth -- either layer-by-layer or step-flow.
Growth mode phase diagrams for similar materials such as STO and LaTiO$_3$ on STO suggest 
that the substrate temperature was not high enough for step flow growth and
that the system should be in the layer-by-layer regime \cite{song_02, ohtomo_02}.
The intensity oscillations, however, are Kiessig (or thickness oscillations)
rather than roughness oscillations.
Each oscillation corresponds to the addition of two rather than 
one unit cell of thickness.
In the present case,
roughness oscillations are likely to be difficult to
observe for two reasons.
First, the illuminated area on the sample is relatively large (4mm by 10mm)
and includes portions of the substrate not in the center of the plume of ablated
material.
Consequently the growth rate varies across the illuminated area, causing the
scattering from different
regions of the film to oscillate at different frequencies.
Second, the substrates did not receive any special treatment ({\it e.g.}\/, etching)
to prepare large, atomically flat terraces.

\begin{figure}[p]
\begin{center}
\includegraphics[width=\linewidth]{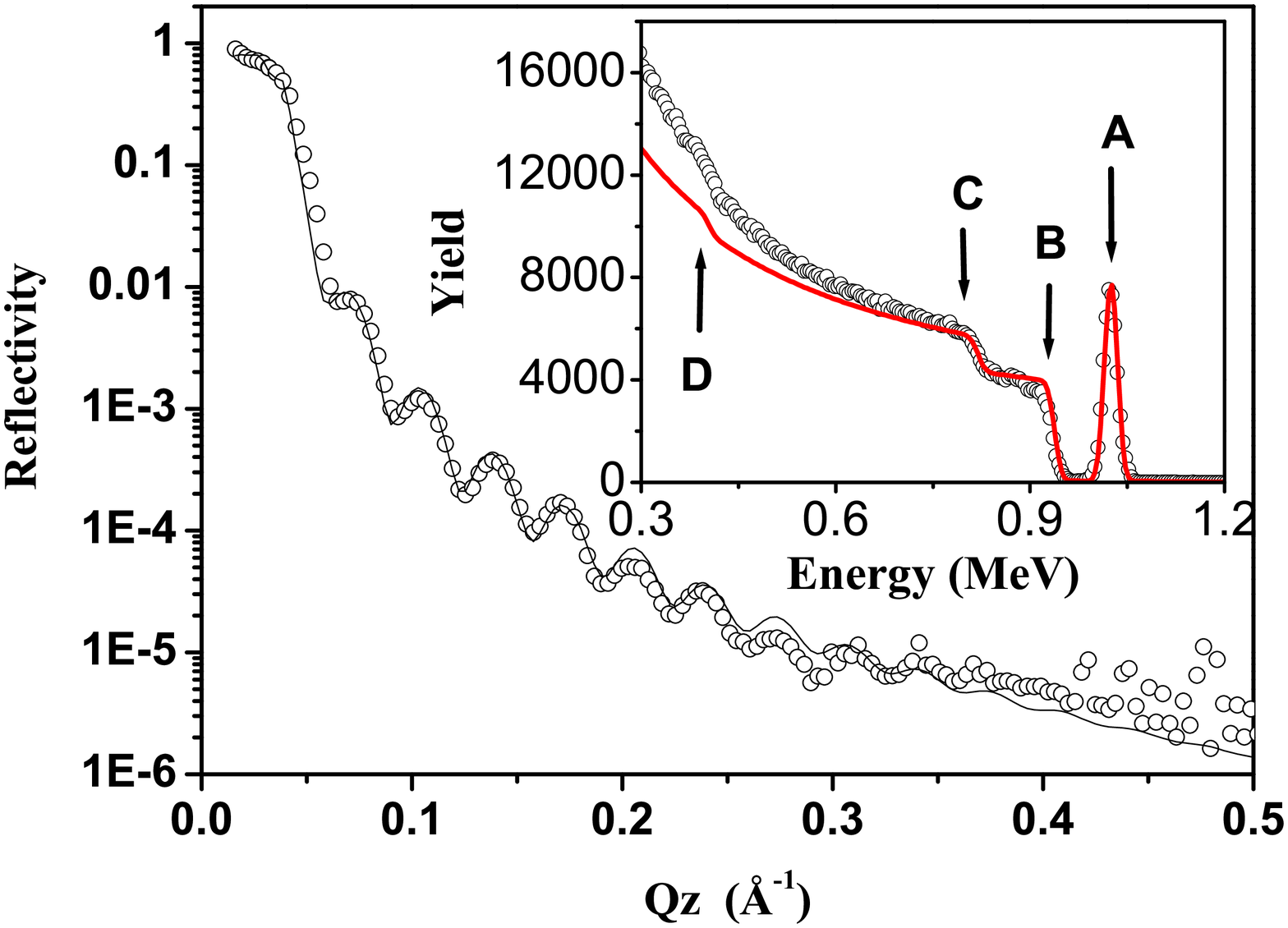}
\end{center}
\caption{Absolute specular reflectivity of EuTiO$_3$ film on SrTiO$_3$\@.
Each circle represents the area under the specular peak in a transverse scan.
The solid line is the best fit to a Parratt line-shape.
Inset: RBS scan of the same film.
Circles are the data and the solid line is the model calculation.
The arrows indicate features in the RBS lineshape associated with
A: europium, B: strontium, C: titanium, and D: oxygen.
}
\label{fig:abs_reflectivity}
\end{figure}

Figure [\ref{fig:abs_reflectivity}] shows the absolute x-ray (specular) reflectivity
as a function of scattered wave vector for
the same film shown in Figure [\ref{fig:oscillations}].
These data were obtained {\it ex situ} using a conventional rotating anode-based
Cu K$_{\alpha}$ x-ray source.
The Kiessig fringes are clearly evident.
The circles are the data and correspond to the integrated intensity obtained from
a rocking scan taken at each $q$ point and then normalized to the incident beam
intensity to obtain the
absolute x-ray reflectivity\cite{gibbs_88}.
The solid line is the best fit
to the Parratt theory using the freeware package {\it Parratt32}\/\cite{als-nielsen}.
The best fit value for the film thickness is $183.9 \pm 9.5\:$\AA\ with a
RMS roughness of $4.7 \pm 1.5\:$\AA\ 
(the error bar represents a doubling of $\chi^2$).
The inset to Figure [\ref{fig:abs_reflectivity}] shows Rutherford Back Scattering (RBS)
data taken on the same sample using singly charged helium ions at
$1.165 \,$MeV\@.
Again, the circles are the data and the solid line is the best fit,
which yielded a best
fit value for the film thickness of $176\:$\AA\@.
The arrows indicate features in the RBS lineshape associated with
A: europium, B: strontium, C: titanium, and D: oxygen.

\begin{figure}[p]
\begin{center}
\includegraphics[width=\linewidth]{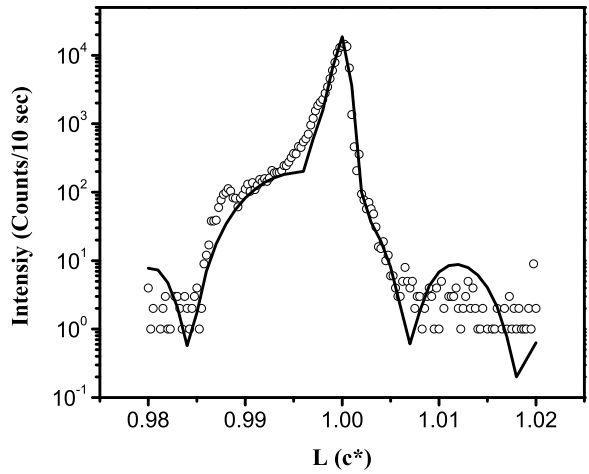}
\end{center}
\caption{
Rod scan through (001) Bragg peak of SrTiO$_3$ and
finite-thickness broadened thin-film ``Bragg'' peak.
Solid line is a simple model calculation consisting of a resolution limited
STO Bragg peak and a simple finite-size line-shape for the ETO film.
}
\label{fig:rod_scan}
\end{figure}

Figure [\ref{fig:rod_scan}] shows an x-ray scattering
scan on the same sample
in the $(0 0 \ell)$ direction in the vicinity
of the $(0 0 1)$ STO Bragg peak.
The solid line is a simple model consisting of a
resolution limited Bragg peak for the STO substrate and
a finite-size line-shape for the ETO thin-film.
Here we have assumed that the thin-film is 47 layers
(183.5\AA\/) thick and that
the ETO reciprocal lattice constant is $0.996 \, a^*$\@,
where $a^*$ is the STO reciprocal lattice constant.
The finite-size model describes the data quite well and
the lattice mismatch in the out-of-plane direction is clearly visible in the raw data.

\begin{figure}[p]
\begin{center}
\includegraphics[width=\linewidth]{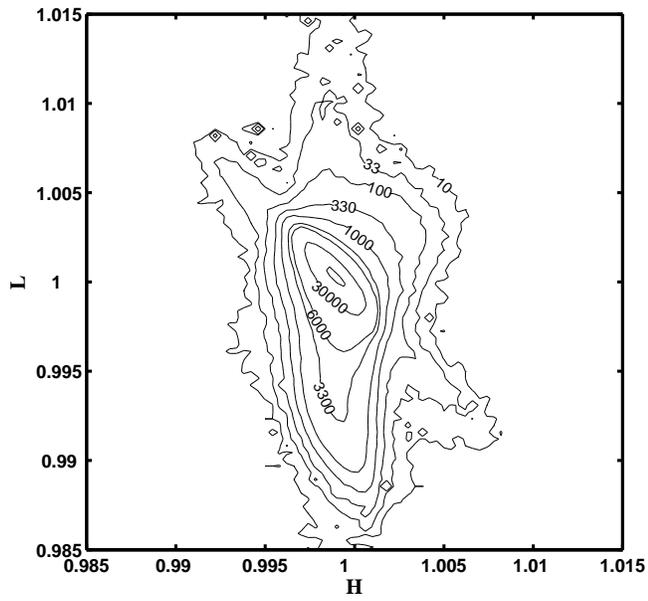}
\end{center}
\caption{
Scattered intensity in the
H-L plane near the (101) Bragg peak of SrTiO$_3$.
Lines are contours of constant scattered intensity.
ETO Bragg peak is roughly at $(1 \; 0 \; 0.996)$.
}
\label{fig:HL-Plane}
\end{figure}

\begin{figure}[p]
\begin{center}
\includegraphics[width=\linewidth]{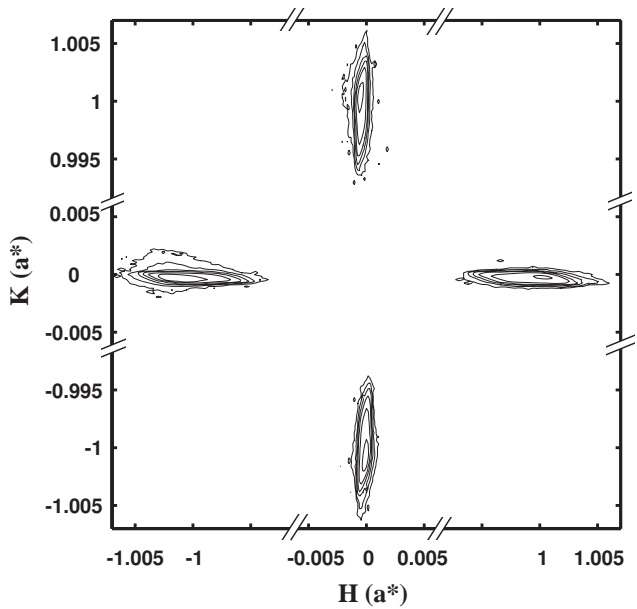}
\end{center}
\caption{
Scattered intensity in the
H-K plane
of SrTiO$_3$
at $\ell = 0.996$.
Lines are contours of constant scattered intensity.
}
\label{fig:HK-Plane}
\end{figure}

To characterize the crystal quality of the ETO thin-film, we performed detailed two-dimensional
maps of the scattered x-ray intensity in the ($h$ $0$ $\ell$) plane near the (1 0 1)
Bragg peak of STO and in the ($h$ $k$ $0.996$) plane.
The first data set is shown in
Figure [\ref{fig:HL-Plane}]\@.
The data is plotted as a contour map with logarithmically spaced
contours.
This data demonstrates that
the in-plane lattice constant of the ETO film is locked to that of the STO substrate.
The streak from ($1\,0\,1$) towards smaller values of $\ell$ indicates
that, consistent with the data in Figure [\ref{fig:rod_scan}] taken near the
($0\,0\,1$) Bragg peak,
the out-of-plane ETO reciprocal lattice constant is roughly 0.4\% smaller than that of the
STO substrate.
The smearing of the data in the $( 1 \, 0 \, \overline{1}) $ direction is due to the resolution
function of the diffractometer.
The two-dimensional data set shown in Figure [\ref{fig:HK-Plane}] 
also shows lines of constant scattered intensity but this time
in the H-K plane at $\ell = 0.996$\@.
The symmetry and location of the four ETO $\langle 1 \, 0 \, 1 \rangle$ Bragg peaks
clearly demonstrate the psuedomorphic growth of the ETO film.

%Figure [\ref{fig:STM_Image}] shows an AFM image of the surface of the same
%ETO film.
%The selected image shows a relatively smooth area of the film between
%large clusters.
%The measured root-mean-square surface roughness is $2.16\:$\AA,
%confirming that the surface of the ETO film is very smooth.

To explore the roles of substrate temperature and laser pulse repetition rate, a series
of growths were performed as the temperature and laser repetition rate were
systematically varied.
Varying the repetition rate from 0.1 to 10Hz
with the substrate temperature fixed at $600^{\circ}$C did not significantly affect the
growth mode.
Similarly,
varying the substrate temperature
from $600^{\circ}$ to $50^{\circ}$C while growing at 0.2Hz did not affect the growth
oscillations.

\section*{Discussion}
In contrast to
many of the rare-earth titanates RTiO$_3$ (R = La, Ce, Nd, Sm, Y),
which exhibit trivalent R and Ti,
EuTiO$_3$ exhibits a divalent Eu
%because of the stable [Xe]$\,4f^7$ electronic configuration
and consequently a tetravalent Ti.
%with the [Ar]$\,3d^0$ electronic configuration.
In this sense
ETO is very similar to the familiar tetravalent perovskite titanates
$A$TiO$_3$ ($A$ = Ba, Sr, Ca).
Therefore, in marked contrast to the LaTiO$_x$ on STO system
where the oxygen partial pressure had to be precisely tuned
to the extremely reducing conditions required to stabilize Ti$^{3+} $\cite{ohtomo_02},
the conditions required to
deposit stoichiometric EuTiO$_3$ thin-films are relatively forgiving.
Combined with the observed structural insensitivity to 
oxygen deficiency \cite{chien_74, mccarthy_69},
ETO is a very promising system for future studies of the oxidation of perovskite titanate thin-films \cite{zhu_99,zhu_01}.

The characterization measurements, both {\it in situ}\/ and {\it ex situ}\/, of the ETO
films were very successful.
First, the Kiessig fringe analysis of the specular reflectivity and the
finite-size analysis
of the (001) Bragg peak give consistent results for the film thickness.
Next, the x-ray (Kiessig) and RBS film thickness measurements agree within
experimental uncertainties.
Finally, the x-ray measurements clearly demonstrate the smoothness of the film.

%The apparent discrepancy in the
%RMS surface roughness obtained from the x-ray measurements and that
%obtained from the AFM measurements is an artifact of the widely different in-plane
%resolutions of the two techniques.
%Although
%both techniques are sensitive to single unit cell height fluctuations,
%the lateral resolution of the AFM is several tens of nanometers.
%Therefore, it is not sensitive to height fluctuations which vary on shorter length
%scales.
%In contrast, the x-ray measurement is sensitive to surface height variations with
%wavelengths comparable to the x-ray wavelength (angstroms).
%Thus, the apparent discrepancy in the surface roughness is consistent with
%single unit cell high features
%(pits or islands) on the surface with lateral dimensions less than roughly
%30 nanometers.

The data clearly demonstrate that,
in the thickness range studied here,
ETO thin films grow epitaxially and pseudomorphically in registry with the
substrate.
The in-plane lattice constant is observed to be locked to that of the STO substrate,
while the out-of-plane lattice constant exhibits a small but measurable strain
of approximately 0.4\%.
If one assumes that the distortion of the ETO unit cell is volume conserving,
then
the lattice mismatch between ETO and STO is roughly 0.2\%.
However, the following caveat needs to be kept in mind.
Based on the growth conditions employed and information available in the literature,
our ETO films are expected be oxygen deficient.
And,
one naively expects that an oxygen deficient sample should have a slightly
dilated real space lattice constant due to additional electrons on the Ti sites.
While the published literature does not report such an effect,
the very small ($\sim$0.2\%) strain we observe is not excluded
by their data.
Since, we have not measured the oxygen concentration,
we are
not able to specify exactly the strain state of a {\it stoichiometric}\/ ETO film.

The observed smoothness of the films and the Kiessig oscillations
indicate that the growth mode is two-dimensional.
The published growth phase diagrams for similar materials
and our temperature and repetition rate studies suggest that the particular
growth conditions utilized here are at too low a temperature to be in the step-flow regime.
Therefore, we suggest that these growth conditions put the system
somewhere near the layer-by-layer growth regime.
However, since we were not able to observe the x-ray roughness oscillations, we cannot
unambiguously determine the growth mode.

%On-going work is developing both the x-ray optics required to use
%smaller x-ray beams and the capability to pre-treat and precisely characterize the
%STO substrates prior to growth.
%In particular, the use of the AFM to characterize the surface of
%HF etched surfaces \cite{kawasaki_94} is quite promising.
%Finally,
%the reported insensitivity of the crystal structure to oxygen vacancies has the potential
%of providing the ability to 
%study the influence of oxygen pressure on the growth mechanism
%without complications due to strain effects.
%With this in mind studies of the effects
%of oxygen pressure, and the use of Eu$_2$Ti$_2$O$_7$ as a target are planned.

\section*{Summary}
In summary,
high quality
epitaxial thin-films of EuTiO$_3$ have been grown
on the (001)
surface of SrTiO$_3$ using 
PLD\@.
{\it In situ}\/ x-ray reflectivity measurements
reveal that the growth is two-dimensional and enable 
real-time monitoring of the film thickness and roughness during growth.
The film thickness, surface mosaic, surface roughness, and strain were characterized
in detail using
{\it ex situ}\/ x-ray diffraction.
The thickness and composition were confirmed with Rutherford Back-Scattering.
The EuTiO$_3$ thin-films grow two-dimensionally, epitaxially,
pseudomorphically,
with no measurable in-plane lattice
mismatch.
The measured out-of-plane lattice mismatch is approximately 0.4\%,

\begin{acknowledgments}
We gratefully acknowledge the assistance of Frank DiSalvo, Dong Park, and John
Sinnott during the target synthesis.
This work was supported by the Cornell Center for Materials Research (CCMR) which is supported by
the National Science Foundation under award DMR~0079992\@.
A portion of this
this work is based upon research conducted at the Cornell High Energy Synchrotron Source
(CHESS) which is supported by the National Science Foundation and the National
Institutes of Health/National Institute of General Medical Sciences under
award DMR~9713424\@.
The construction of the G-line facility was partially
supported by the National Science Foundation
under awards DMR 9970838 and DMR 0114094.
\end{acknowledgments}

\bibliography{WangReferences}% Produces the bibliography via BibTeX.

\end{document}